\documentclass[aps,pre,twocolumn,showpacs,superscriptaddress]{revtex4-1}
\usepackage{graphicx}
\usepackage{xfrac}
\usepackage{color}
\begin{document}
\title{Identifying transitions in finite systems by means of partition
function zeros and microcanonical inflection-point analysis: A comparison for
elastic flexible polymers}
\author{Julio C.\ S.\ Rocha}
\email[E-mail: ]{jcsrocha@physast.uga.edu}
\affiliation{Center for Simulational Physics, The
University of Georgia, Athens, Georgia 
30602, USA}
\author{Stefan Schnabel}
\email[E-mail: ]{stefan.schnabel@itp.uni-leipzig.de}
\affiliation{Institut f\"ur Theoretische Physik and Centre for Theoretical
Sciences (NTZ), 
Universit\"at Leipzig, Postfach 100920, D-04009 Leipzig, Germany}
\author{David P. Landau}
\email[E-mail: ]{dlandau@hal.physast.uga.edu}
\homepage[\\ Homepage: ]{http://www.csp.uga.edu}
\affiliation{Center for Simulational Physics, The University of Georgia,
Athens, Georgia 
30602, USA}
\author{Michael Bachmann}
\email[E-mail: ]{bachmann@smsyslab.org}
\homepage[\\ Homepage: ]{http://www.smsyslab.org}
\affiliation{Center for Simulational Physics, The University of Georgia,
Athens, Georgia 
30602, USA}
\affiliation{Instituto de F\'{\i}sica,
Universidade Federal de Mato Grosso, Cuiab\'a (MT), Brazil}
\affiliation{Departamento de F\'{\i}sica,
Universidade Federal de Minas Gerais, Belo Horizonte (MG), Brazil}
\date{\today}
\begin{abstract}
For the estimation of transition points of finite elastic,
flexible polymers with chain lengths from $13$ to $309$ monomers,
we compare systematically transition temperatures obtained by the Fisher
partition function zeros approach with recent results from microcanonical
inflection-point analysis. These methods rely on accurate numerical estimates
of the density of states, which have been 
obtained by advanced multicanonical Monte Carlo sampling techniques. 
Both the Fisher zeros method and microcanonical inflection-point analysis
yield very similar results and enable the unique 
identification of transition points in finite systems, which is typically
impossible in the conventional canonical analysis of thermodynamic
quantities.
\end{abstract}
\maketitle
%
\section{Introduction \label{introduction}}
%
Phase transitions are among the most fascinating phenomena in nature and 
huge efforts have been made to understand the features that
characterize these cooperative processes for many different systems in a
general and systematic way. 
Strictly speaking, thermodynamic phase transitions occur only in the
thermodynamic
limit, i.e., for infinitely large systems. However, recent growing 
interest has also involved finite systems. Prominent representatives for such
systems are finite polymer chains and, in particular, proteins. 
Because of surprisingly manifest common properties of transitions in finite
and infinite systems, the question arose to what extent the relationship
between
``pseudo-transitions'' in finite systems and their infinite-system
counterparts can be stressed. It is well known that the precise determination
of the location of transitions in finite systems is typically ambiguous and
different fluctuating quantities suggest different
points in parameter space as transition points. In the thermodynamic limit,
scale freedom would let this space collapse to a single unique transition
point. However, most contemporary problems in soft condensed matter and
technology are apparently of small size, for which the thermodynamic limit is
not applicable at all. For this reason, it is necessary to verify if
the methods of statistical analysis that have been developed for
infinitely large systems and have proven to be so extremely successful in
these cases can be employed for, or adapted to, finite systems as well. 

Another important aspect is the fact that computer simulations open a
completely new view on statistical physics, as only the most recently
developed computational methods and algorithms enable the accurate study of
fundamental statistical quantities that could hardly be approached by
theoretical methods in the course of the establishment of the theory of
complex phenomena and phase transitions in the past decades. One such
quantity is the density of states $g(E)$, i.e., the number of system
configurations within a given energy interval. Its logarithm can be associated
with the entropy of the system in energy space, $S(E)=k_\mathrm{B}\ln\,g(E)$,
and the first derivative with respect to energy yields the inverse
temperature $\beta(E)=dS(E)/dE$. It has been shown recently that the
careful analysis of inflection points of this quantity reveals all
transitions in the system uniquely and without any ambiguity~\cite{sslb1}.
Since in this approach
the temperature is considered to be a derived quantity and a function of
energy, this method is a representative of microcanonical statistical
analysis.

In this paper, we will also make use of the density of states, but we are
going to interpret its features in a canonical way by considering the
partition function $Z(T)$ of the system as a function of the (canonical)
temperature $T$. The thermodynamic potential associated with the canonical
ensemble (we consider fixed system size $N$ and volume $V$) is the free
energy $F(T)=-k_\mathrm{B}T\ln\,Z(T)$. Thermodynamic phase transitions are
located in temperature space, where a derivative of $F$ of a certain
order exhibits a singularity~\cite{ehrenfest1,%
ldlandau1,fisher1,kadanoff1,heller1}. Examples are the canonical entropy
$S(T)=-(dF(T)/dT)_{N,V}$ and response quantities such as the heat capacity
$C_V=T(dS(T)/dT)_{N,V}=-T(d^2F(T)/dT^2)_{N,V}$.
Yang and Lee were the first to relate catastrophic singularities to 
partition function zeros in the grand canonical
ensemble by introducing complex fugacities~\cite{yanglee1}.
Fisher evolved this idea for the canonical partition function by introducing
a complex temperature plane~\cite{fisher2}. 

There is extensive literature on applications of such methods to various
physical systems such as spin models (see, e.g.,
Refs.~\cite{jankekenna1,hwang1,fonseca1}), proteins~\cite{hansmann1,wang1},
and to polymers~\cite{taylor1,hu1}. Most applications of the
partition function zero analysis method are considered to be alternative
approaches to scaling properties near phase transitions in large systems. 
However, this method is also promising for the identification
and characterization of analogs of phase transitions in finite systems, in
particular in finite linear polymer chains that are known to exhibit a variety
of structural transitions which sensitively depend on the chain
length~\cite{sslb1,svbj1,sbj1,seaton1}. The understanding of these structure
formation processes
is relevant from both fundamental scientific and applied technological
perspectives of molecular building-block systems.

Typically these processes are accompanied by nucleation transitions, where 
crystalline shapes form from a liquid or vapor phase.
Crystalline or glass-like structures of single polymer chains can serve as the
basic elements of larger
assemblies on nanoscopic scales; and beyond that, the crystallization behavior
exhibits strong similarities
to the cluster formation of colloidal (or atomic) particles~\cite{sbj1}.
The nucleation is governed by finite-size and surface effects, where
functionalization is based on the individual structural properties of small
molecules forming large-scale composites~\cite{sbj1}. 
These effects can be analyzed by means of microcanonical
thermodynamics~\cite{gross1}, in which case
transition properties can be derived directly and systematically from
the caloric entropy curve~\cite{sslb1}. This approach has been successfully
applied to a variety of structural transitions in macromolecular
systems such as folding~\cite{chen1,taylor2,mb1,sslb1},
aggregation~\cite{jbj1}, and adsorption processes of polymers and
proteins~\cite{liang1,mjb1}. One particular problem that has gained increased
interest recently is the influence of the interaction range on the stability
of structural phases~\cite{taylor2,gnvb1}. This has been addressed by means of
systematic microcanonical analyses in discrete and continuous polymer models.

In principle, once the density of states $g(E)$ is given, the partition
function can easily be calculated and its zeros identified. However, examples
of systems for which $g(E)$ can be calculated exactly, or quite accurately
by theoretical methods, are very rare. It requires sophisticated
numerical methods such as generalized-ensemble Monte Carlo sampling that allow
for accurate estimates of $g(E)$. Among the most popular methods are
multicanonical sampling~\cite{muca1,muca2} and the Wang-Landau
method~\cite{wl1}. These methods are capable of scanning the entire phase
space effectively in a single simulation.

Compared to recent studies on partition function zero analyses of 
polymers such as Ref.~\cite{taylor1}, we here employ a more realistic 
coarse-grained model for elastic, flexible polymers with continuous, 
distance-dependent monomer-monomer interactions based on van der Waals forces. 
Recently developed sophisticated simulation methodologies 
specific to this model~\cite{sjb1} enable a very precise estimation of 
fundamental statistical quantities such as the density of states. This is 
essential for the careful identification of low-entropy phases that include 
liquid-solid and solid-solid transitions. For finite systems, these transitions 
are strongly affected by finite-size effects, which are of particular interest 
in this comparative study of advanced statistical analysis methods. One major 
question is whether the partition function zeros method, which is effectively a 
canonical approach, is capable of revealing the same intricate details of these 
effects as the microcanonical inflection-point analysis~\cite{sslb1}. 
For this purpose, we systematically analyze the canonical partition function 
zeros for all
chain lengths ranging from $13$ to $309$ monomers in this model and
identify and classify all structural transitions. Since the finite-size effects 
in the solid phases are surface effects specific to the explicit chain length, 
transitions in between them do not exhibit obvious scaling 
properties~\cite{vbj1,sbj1,sslb1}. Therefore, scaling considerations are not in 
the focus of this study.

This paper is organized as follows: In Sec.~\ref{sec:I}, we review  
the partition function zeros approach and describe the numerical methods
used for the estimation of the density of states and for the identification
of the Fisher zeros. This section also includes a brief discussion of the
microcanonical inflection-point analysis.
The results of our study are
presented in Sec.~\ref{results}, where we first discuss 
the different scenarios in the liquid-solid and solid-solid transition 
regimes thoroughly by investigating the zero maps for four representative 
examples that differ in the processes of Mackay and anti-Mackay overlayer 
formation. We then generalize and summarize the results obtained by the 
zeros method for all polymers with chain lengths up to 309 monomers and compare 
with former results obtained by microcanonical inflection-point 
analysis~\cite{sslb1}.
The paper is concluded by a summary in
Sec.~\ref{summary}.
%
\section{Methods and Model \label{methods}}
\label{sec:I}
\subsection{Partition function zeros and thermodynamics \label{deduction}}
%
We consider a polymer system in thermal equilibrium with a heat bath that is
described
by the canonical $NVT$ ensemble (constant particle number $N$, volume $V$,
and temperature $T$). This ensemble connects microscopic quantities and
thermodynamical properties via
statistical relations described by the canonical partition function $Z$.
In thermal equilibrium, the probability for a discrete energetic state is
$p_m = g_m e^{-\beta E_m}/Z$,
where $g_m$ denotes the density of states at each energy $E_m$;
$\beta = 1/k_\mathrm{B}T$ is the inverse thermal energy and $k_\mathrm{B}$ is
the Boltzmann constant.
In this work the units are chosen so that $k_\mathrm{B} = 1$. 
For a discrete ensemble of energetic states, the partition function reads
\begin{equation}
Z = \sum_m g_me^{-\beta E_m} = e^{-\beta E_0}\sum_m g_me^{-\beta (E_m - E_0)},
\label{partition_function}
\end{equation}
where we have extracted the Boltzmann factor of the ground state for future
convenience.
All essential thermodynamic quantities such as entropy and response
functions like the heat capacity derive from the free energy $F =
-\ln{Z}/\beta$.
\begin{figure}
    \includegraphics[height=5.5cm,width=8.5cm]{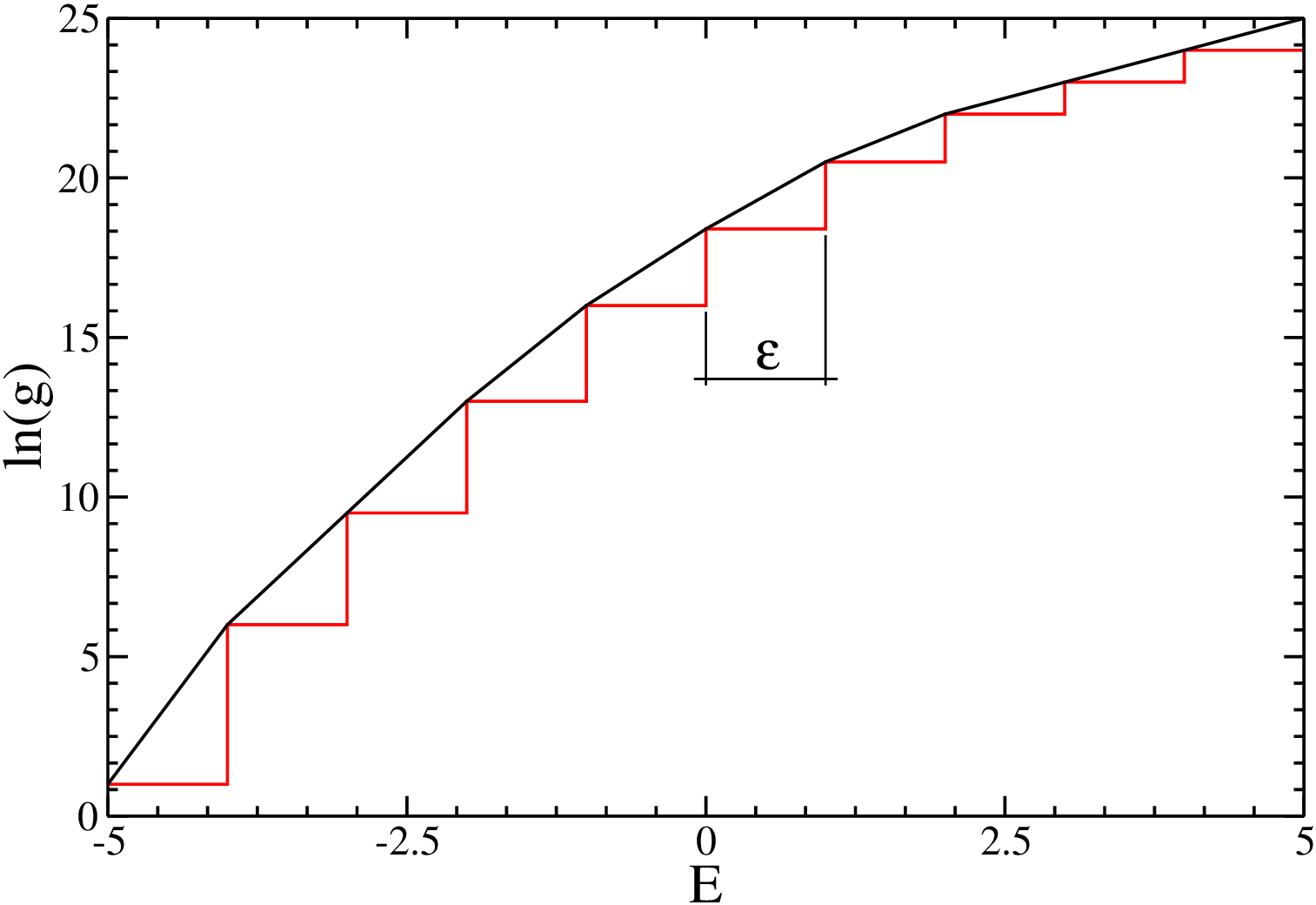}
  \caption{Pictorial demonstration of the discretization of a continuous
density  of states over an energy range which is divided in $n$ bins of size
$\varepsilon$.
Here the bins are labeled from $0$ to $n-1$, thus the energy of the $m$th 
bin is $E_m = E_{0} + m\varepsilon$.
All states with energy between $E_m$ and $E_m + \varepsilon$ are recorded in 
the $m$th bin $g_m$.
\label{density}}
\end{figure}

For the subsequent analysis of a model with a continuous energy spectrum,
it is necessary to discretize the density of states. Estimates obtained by
means of generalized-ensemble Monte Carlo methods such as
multicanonical~\cite{muca1,muca2} and Wang-Landau sampling~\cite{wl1} are
naturally discrete in energy space (see Fig.~\ref{density}). If the energy bin
size is chosen to be $\varepsilon$, the partition
function~(\ref{partition_function}) can be rewritten as
\begin{equation}
Z = e^{-\beta E_0} \sum_{m=0}^{n-1} g_m e^{-\beta m \varepsilon},
\end{equation}
where $n$ denotes the total number of bins.

Defining $x \equiv e^{-\beta \varepsilon}$, the partition function
can assume the form of a polynomial
\begin{equation}
Z = e^{-\beta E_0} \sum_{m=0}^{n-1} g_m x^m = e^{-\beta E_0} \prod_{j=1}^{n-1}
(x-x_j).
\label{eq:polyz}
\end{equation}
In the latter expression,
the polynomial was decomposed into linear factors  $(x-x_j)$, where $x_j$
denotes the $j$th 
zero (or root) of the polynomial. 
With the polynomial defined in this way, the density of states can cover the
entire space 
of energy for both positive and negative energies. 
Note that $x\ge 0$; if $T \rightarrow 0$ then $x \rightarrow 0$,
whereas $x \rightarrow 1$, if
$T \rightarrow \infty$.

In Eq.~(\ref{eq:polyz}), $Z$ is written as a polynomial of degree $n-1$ which
has $n-1$, generally complex, roots. Since $Z\in \Re$ and for a finite system
always $Z>0$
and since the coefficients $g_m$ are
nonzero positive real numbers, the roots must occur as complex conjugate pairs
$a_j \pm i b_j$ with $a, b \in \Re$. Real-valued roots must be
negative. 

Once the partition function is determined thermodynamic
quantities can be extracted from the
the Helmholtz free energy $F$. 
The internal energy is
\begin{equation}
U = \langle E \rangle = -\frac{\partial \ln{Z}}{\partial \beta}
\end{equation}
and, most interesting for the following consideration, the specific heat at
constant volume reads
\begin{equation}
c_V =\frac{1}{N}\left(\frac{\partial U}{\partial T}\right)_V=
\frac{k_\mathrm{B}\beta^2}{N} \frac{\partial^2 \ln{Z}}{\partial \beta^2}.
\end{equation}
Inserting the factorization~(\ref{eq:polyz}), these quantities can also be
expressed by the Fisher zero components:
\begin{equation}
U =  E_0  + \sum_{j=1}^{n-1} \left( \frac{\varepsilon x}{x-x_j} \right) =
E_0 + \sum_{j=1}^{n-1} \left( \frac{\varepsilon
x(x-a_j)}{(x-a_j)^2+b_j^2}\right),
\end{equation}
and
\begin{eqnarray}  \nonumber
&&\hspace*{-5mm}c_V = \frac{k_\mathrm{B}x(\ln{x})^2}{N} \sum_{j=1}^{n-1}
\left( \frac{-x_j}{(x-x_j)^2}\right) \\ 
&& = \frac{k_\mathrm{B}x(\ln{x})^2}{N} \sum_{j=1}^{n-1} \left(
\frac{-a_j(x-a_j)^2 +b_j^2(2x-a_j)}{[(x-a_j)^2+b_j^2]^2}\right).
\label{cv_zero}
\end{eqnarray}
Obviously, this expression can only become singular at $x=a_j$, if
$b_j=0$, i.e., if the $j$th zero lies on the positive real axis. 
According to Yang and Lee, zeros that come arbitrarily close to the real
axis in the thermodynamic limit mark the transition points. This is essential
for our 
study as we are interested here exclusively in transition properties of
polymers of finite length. Therefore, we do not expect to find any real-valued
zeros in the analysis of the complex-zero space of these systems. Rather, we
will identify the zeros closest to the positive real axis which are called
the leading zeros because they contribute most to the quantity of interest,
if $x\approx a_j$. If such zeros have a rather isolated appearance in the
distribution of the zeros in the complex map near the positive real axis,
they represent a signal in that quantity that might become a singularity
in the infinitely large system. At least, in the finite system, they indicate
increased thermal activity. Canonical quantities such as the specific
heat typically possess a peak or a ``shoulder'' in those regions in
temperature space. 

Technically, apart from finite-size scaling, there are two possibilities to
define transition points for finite systems by means of partition function
zeros. Either one considers the zero as if it lies on a circle (in
first-order like transitions, the transition-state zeros distribute indeed
near a circular line), in which
case the radius defined via $|x_j|^2=a_j^2+b_j^2$ can be used to locate the
intersection point on the positive real axis: $x_\mathrm{c}\equiv a_j'=|x_j|$.
Alternatively, since $b_j$ will be small near the positive real axis, one can
also simply choose $x_\mathrm{c}=a_j\approx |x_j|$. Either way, by performing
the projection upon the real axis, a specific-heat singularity is
mimicked \emph{even for a finite system}. The transition point can then be
defined by
\begin{equation}
 T_\mathrm{c} = -\frac{\varepsilon}{k_\mathrm{B}\ln{|x_j|}}.
\label{transition_temperature}
\end{equation}
On this basis, conclusions
about the structural transitions of finite-length flexible polymers will be
drawn in this study, but these transitions should
not be confused with the strictly defined thermodynamic phase transitions in
the Yang-Lee sense.

The accurate estimation of the partition function zeros requires two separate
parts that for a complex system can only be accomplished computationally.
First, generalized-ensemble Monte Carlo simulations have to be performed to
obtain the density of states. Second, all zeros of the polynomial form of the
partition function must be identified. Since a polynomial of degree five or
higher has no algebraic solution in general,
as stated by the Abel-Ruffini theorem, the zeros can only be found
by means of numerical computation. We will review the polymer model and
the simulation and analysis methods used in the following. 
%
\subsection{Coarse-grained polymer model \label{model}}
%
A linear polymer of length $L$ is formed by concatenation of 
$L$ identical chemical units called monomers.
Each monomer is composed of several atoms, thus the size
of the chain suitable
for simulation is limited by the computational resources and methods currently
available.
For the study of generic thermodynamic properties of polymers, however, 
all-atom models can typically be replaced  
by a simpler coarse-grained representation with effective interactions.
We here consider such a generic coarse-grained model for linear, elastic,
flexible
polymers~\cite{svbj1}.
Non-bonded monomers interact pairwise via a truncated and shifted
Lennard-Jones (LJ) potential
\[
V^\mathrm{mod}_\mathrm{LJ}(r_{ij}) = V_\mathrm{LJ}(\min(r_{ij},r_c)) -
V_\mathrm{LJ}(r_c), 
\]
where $r_{ij}$ denotes the distance between the $i$th and the $j$th
monomer, $r_c$ is the cutoff distance, and
\[
V_\mathrm{LJ}(r) = 4 \epsilon \left[ \left( \frac{\sigma}{r} \right)^{12} -
\left(
\frac{\sigma}{r} \right)^{6} \right]
\]
is the standard LJ potential. In this work the LJ parameters were chosen as
$\epsilon=1$, $\sigma = 2^{-1/6}r_0$, and $r_c = 2.5\sigma$.

The elastic bonds between monomers adjacent along the chain are modeled by the
finitely extensible nonlinear elastic (FENE) potential~\cite{fene1}
\[
V_\mathrm{FENE}(r_{ii+1}) = -\frac{K}{2}R^2 \ln{ \left[ 1 - \left(
\frac{r_{ii+1} - r_0}{R} \right)^2 \right]}.
\]
This potential possesses a minimum at $r_0$ and diverges for $r \rightarrow
r_0 \pm R$. $K$ is a spring constant and we set
the parameters as $R=0.3$, $r_0 = 0.7$, and $K=40$.
%
\subsection{Numerical methods \label{numerical}}
%
\subsubsection{Monte Carlo sampling in a generalized ensemble
\label{simulation}}
Since the simulation of structural phases of polymers is challenging, even for
a coarse-grained model and moderate system sizes, a sophisticated advanced
Monte Carlo update set~\cite{sjb1} was applied in combination with
multicanonical sampling~\cite{muca1,muca2,sjb1}. The majority of moves
consisted of attempted displacements of single monomers within a sphere around
their original location. Depending on energy $E$ and number of monomers $N$
the radii of these spheres were chosen such that high acceptance rates could
be achieved for all energies and system sizes. In addition, we used
bond-rebridging moves, where all monomers keep their position, but the linkage
between them is altered. Furthermore, a novel cut-and-paste move was developed
in which one monomer is removed and reinserted in an entirely different
location within the polymer chain.

Most of the data were produced in a single simulation by sampling a
generalized
``grand-multicanonical'' ensemble~\cite{sjb1}. The main goal was to avoid free
energy barriers by enabling the system to change its size. Therefore, in
addition to the trial update schemes described above, a Monte Carlo step was
introduced by means of which single monomers could randomly be added or
removed. A weight function $W(E,N)$ assured that all energies and sizes were
visited with the same probability. It was tuned using a delayed Wang-Landau
procedure, in which the modification factor of the original
Wang-Landau method is made weight-dependent. If the multicanonical weight
function at Monte Carlo ``time'' $t$ is denoted by $W_t$,
then it is modified after the next update to
\begin{equation}
W_{t+1}(E,N) =W_{t}(E,N)/
f^{W_{t}(E,N)/W_{t-d}(E,N)}
\end{equation}
for $E=E_{t-d}$, $N=N_{t-d}$. For other values of $E$ and $N$, the weight remains unchanged as in a
conventional multicanonical simulation. Therefore, the effect of the 
Wang-Landau modification factor $f$ to smooth out the free-energy landscape is
delayed
by $d$. This slows down the saturation speed of Wang-Landau sampling and
enables a better efficiency in exploring phase space regions of low entropy
at low energy, in particular in isolated regions that might contain hidden
barriers. For the polymer system considered here, this is particularly
relevant in the solid-solid transition regime.
A sufficiently large delay for the polymer model considered here is obtained
by the choice $d=10^4$. 

Once the weights had converged data were generated in a
grand-multicanonical production that consisted of approximately
$2\times10^{12}$ Monte
Carlo moves and consumed about 0.5 CPU years.
%
\subsubsection{Zeros finder \label{finder}}
%
Computing the zeros of polynomials can be posed as an eigenvalue
problem~\cite{dooren1,numrec1}.
Consider the matrix pair ($\mathbf{A}$,$\mathbf{B}$) where
\begin{equation} 
\mathbf{A} = \left[ \begin{array}{cccccc}
    0    &     0    &     0    & \cdots &    0   &  -g_0   \\
    1    &     0    &     0    & \cdots &    0   &  -g_1   \\
    0    &     1    &     0    & \cdots &    0   &  -g_2   \\
    0    &     0    &     1    & \cdots &    0   &  -g_3   \\
 \vdots  &  \vdots  &  \vdots  & \ddots & \vdots & \vdots  \\
    0    &     0    &     0    & \cdots &    1   & -g_{n-1}
\end{array} \right]
\label{companion_matrix}
\end{equation}
is the Frobenius companion matrix related to a monic polynomial~\cite{rem2} of
degree $n$~\cite{linden1}, and 
\begin{equation} 
\mathbf{B} = \left[ \begin{array}{cccccc}
    1    &     0    &     0    & \cdots &     0    &     0   \\
    0    &     1    &     0    & \cdots &     0    &     0   \\
    0    &     0    &     1    & \cdots &     0    &     0   \\
 \vdots  &  \vdots  &  \vdots  & \ddots &  \vdots  &  \vdots \\
    0    &     0    &     0    & \cdots &     1    &     0   \\
    0    &     0    &     0    & \cdots &     0    &  g_{n}
\end{array} \right].
\label{matrix_B}
\end{equation}
Then a straightforward computation shows that
\begin{equation}
\det{(x\mathbf{B} - \mathbf{A})} = \sum_m g_mx^m = P(x).
\label{polynomial_C}
\end{equation}
On the other hand, the well-known generalized eigenvalue problem
(GEP)~\cite{watkins1} can be stated as
\begin{equation}
\det{(\lambda\mathbf{B} - \mathbf{A})} = 0.
\label{GEP}
\end{equation}
By comparing Eqns.~(\ref{polynomial_C})~and~(\ref{GEP}) one
finds that eigenvalues 
of the matrix pencil ($\mathbf{A}$,$\mathbf{B}$) are the zeros of $P$, i.e.,
$x_k=\lambda_k$.
The GEP can be solved by the QZ algorithm~\cite{moler1}, just after
performing a
balance on the matrix pair ($\mathbf{A}$,$\mathbf{B}$), which is very
important for accuracy~\cite{parlett1,ward1,dooren2}.
Both of these algorithms can be found in LAPACK~\cite{lapack}.
Alternatively, as implemented in {\it Mathematica}~\cite{weiss1}, one
can write a companion matrix of $P$ as
\begin{equation}
\mathbf{C} = \left[ \begin{array}{cccccc}
    \sfrac{-g_1}{g_0}    &      \sfrac{-g_2}{g_0}    &   \sfrac{-g_3}{g_0}    & \cdots &     \sfrac{-g_{n-1}}{g_0}   &  \sfrac{-g_n}{g_0}    \\
    1    &     0    &     0    & \cdots &    0   &     0 \\
    0    &     1    &     0    & \cdots &    0   &  0   \\
    0    &     0    &     1    & \cdots &    0   & 0   \\
 \vdots  &  \vdots  &  \vdots  & \ddots & \vdots & \vdots       \\
    0    &     0    &     0    & \cdots &    1   & 0 
\end{array} \right].
\label{companion_matrixB}
\end{equation}
Then the zeros of $P$ are obtained directly by diagonalization of
$\mathbf{C}$
and given by
\begin{equation}
x_k = \frac{1}{\lambda_k}.
\end{equation}
This method is more time consuming but also more robust than the previous
one.

We employed both methods for the estimation of the partition function
zeros~(\ref{eq:polyz}).
%
\subsubsection{Microcanonical inflection-point analysis\label{microcanonical
entropy}} 
%
%
\begin{figure*}
    \begin{tabular}{c c c}
    \includegraphics[height=7.5cm,keepaspectratio=true]{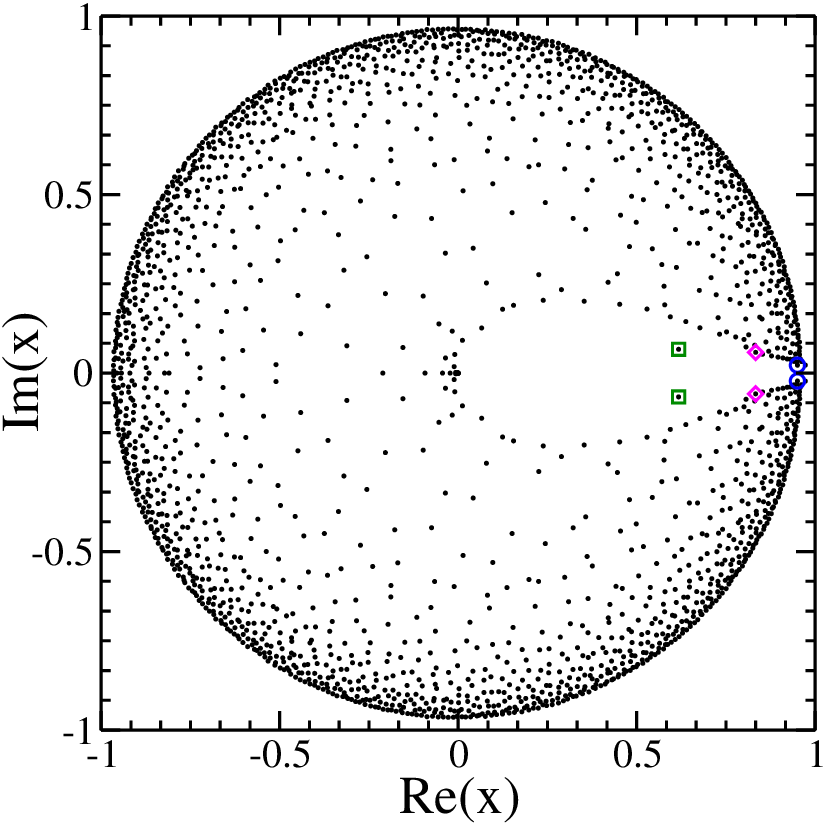} & &
    \includegraphics[height=7.5cm,keepaspectratio=true]{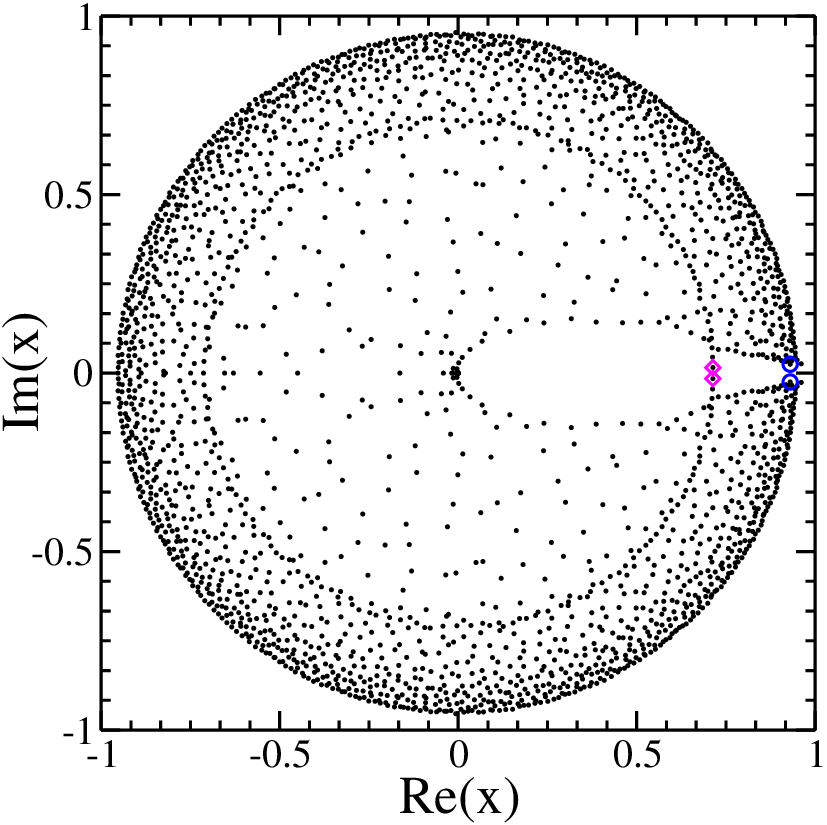} \\ 
\vspace{5mm}
     (a) & & (b) \\
    \includegraphics[height=7.5cm,keepaspectratio=true]{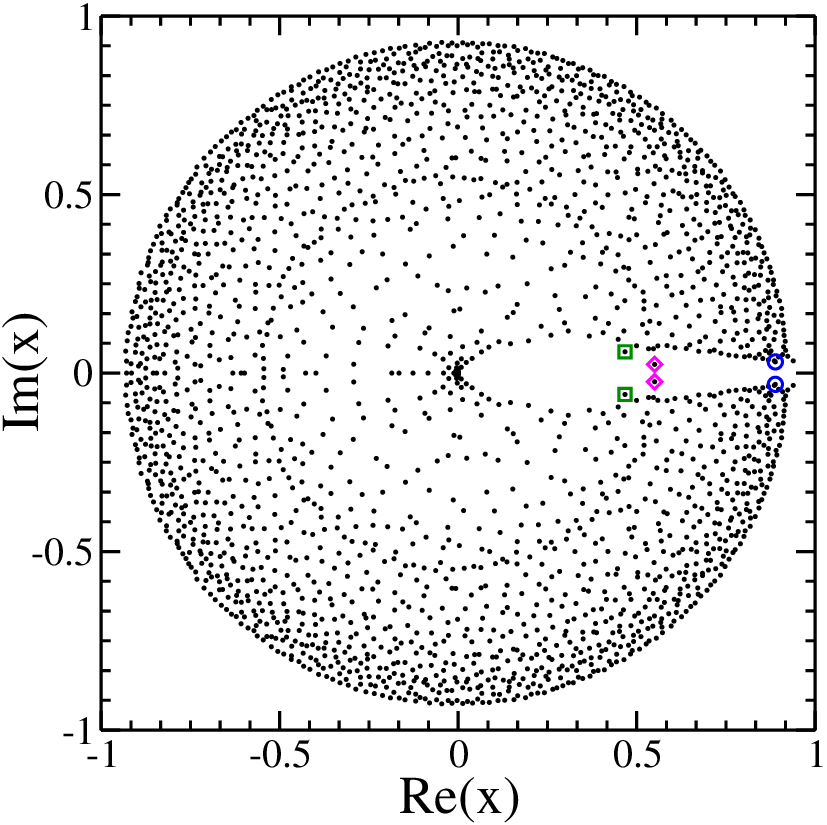} & &
    \includegraphics[height=7.5cm,keepaspectratio=true]{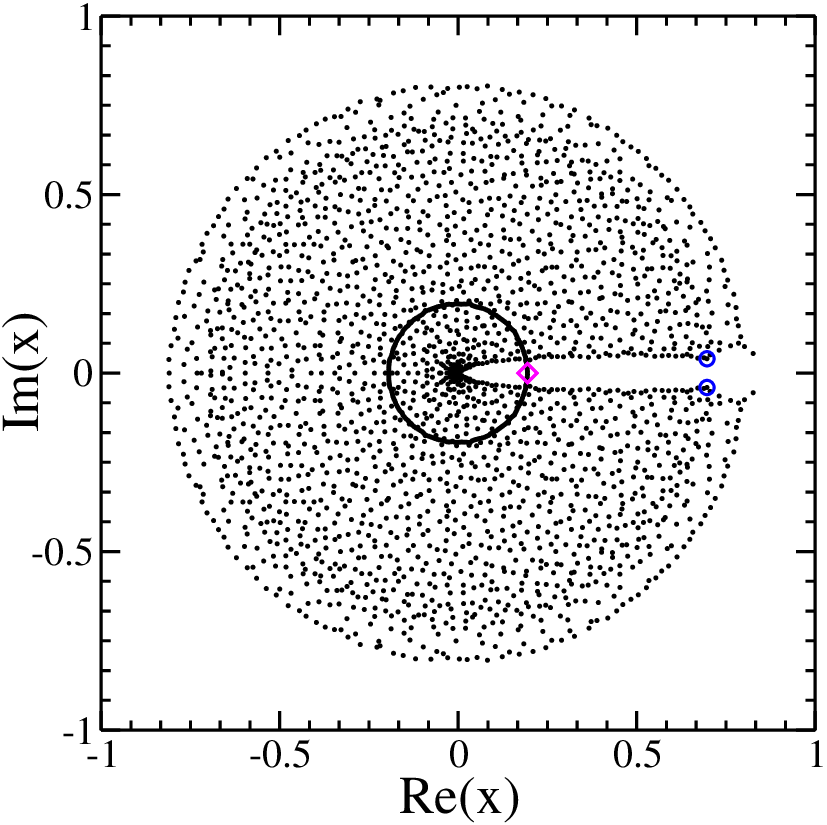} \\
     (c) & & (d)
  \end{tabular}
  \caption{Complex plane map of the partition function zeros for chain size:
(a) $L=35$, (b) $L=55$, (c) $L=90$, and (d) $L=300$.
The leading zeros are highlighted as follows: From $x=0$ to $1$ green
squares
denote ``solid-solid'' transitions, 
magenta diamonds denote ``liquid-solid'' transitions, and blue circles denote
``gas-liquid'' transitions. \label{root_map}}
\end{figure*}
\begin{figure}
\includegraphics[height=5.5cm,width=8.5cm]{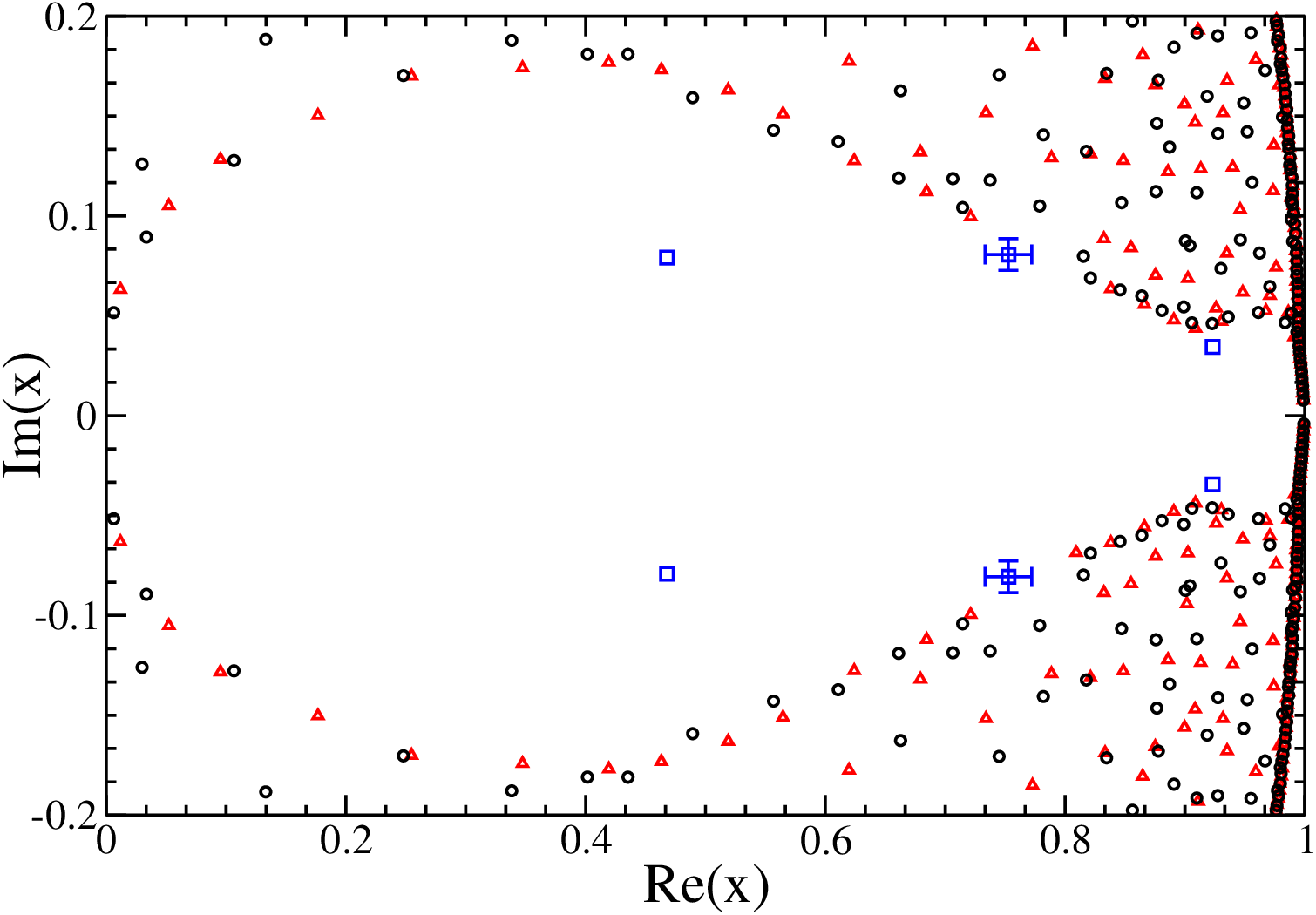}
\caption{Zoom into the zeros map for $L=35$. 
Black circles and red triangles represent the zeros
obtained in 
two different simulations.
Whereas the positions of nonleading zeros vary, the
leading zeros are very close to each other and the overall distribution
pattern is very similar.
The blue squares
represent the
average
values of the leading zeros over ten different simulations. Error bars are
shown for the leading zero that corresponds to the liquid-solid transition;
in the other cases the error is  
smaller
than the symbol size.  \label{error_zoom}}
\end{figure}
\begin{figure*}
    \begin{tabular}{c c c}
    \includegraphics[height=5.5cm,width=8.5cm]{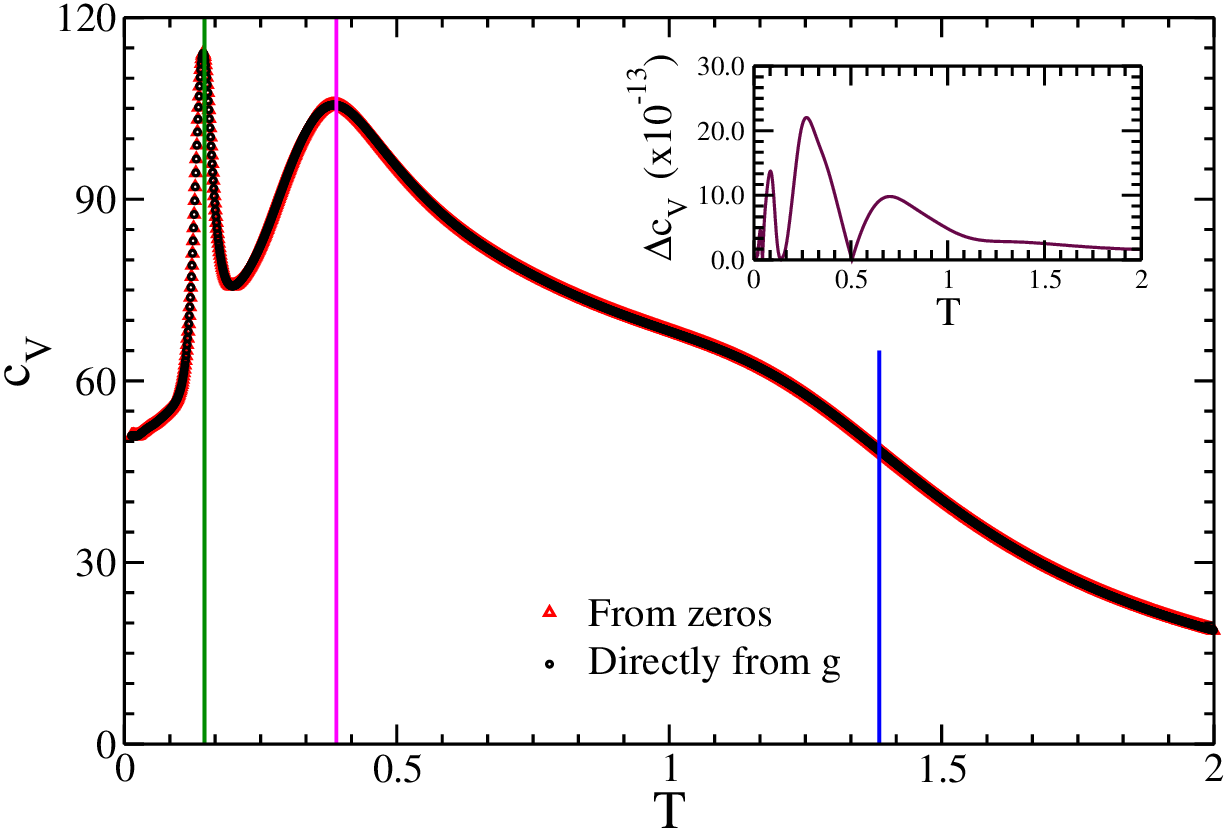}
& &
    \includegraphics[height=5.5cm,width=8.5cm]{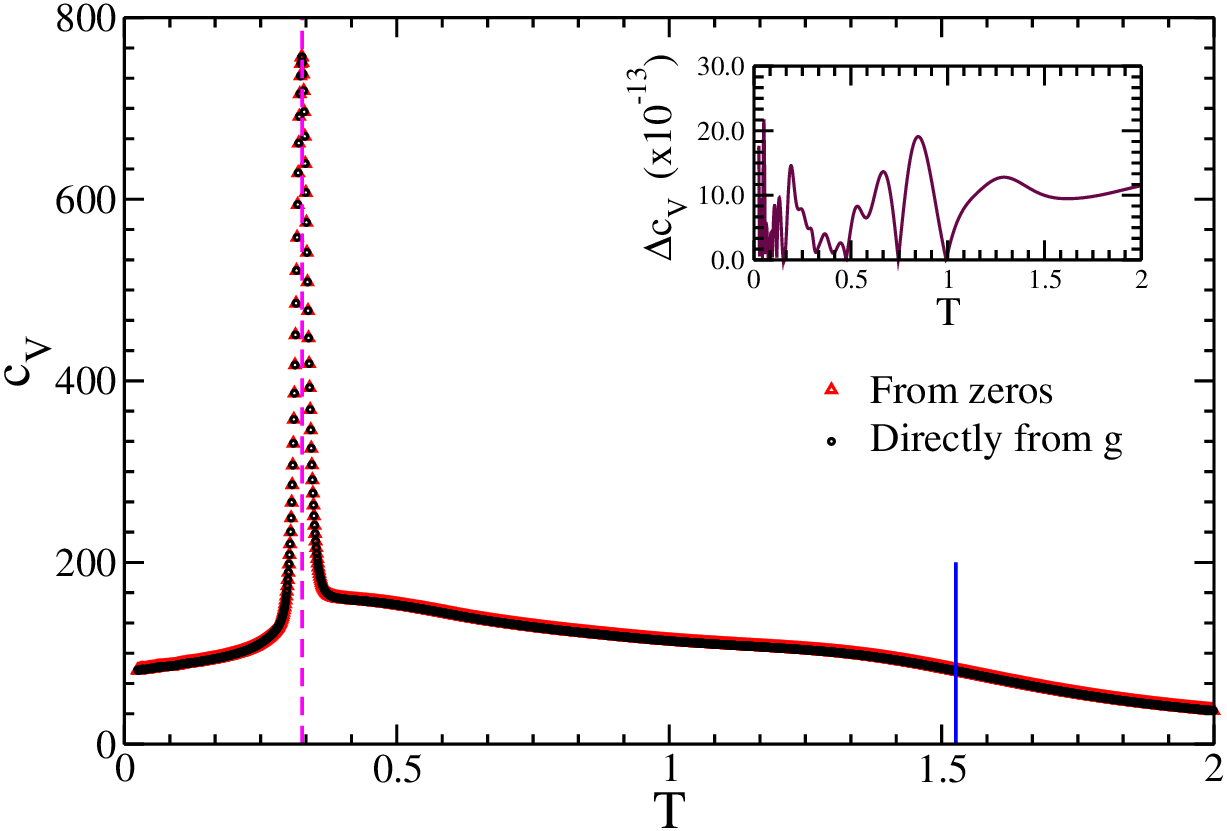}
\\ \vspace{1cm}
     (a) & & (b) \\
    \includegraphics[height=5.5cm,width=8.5cm]{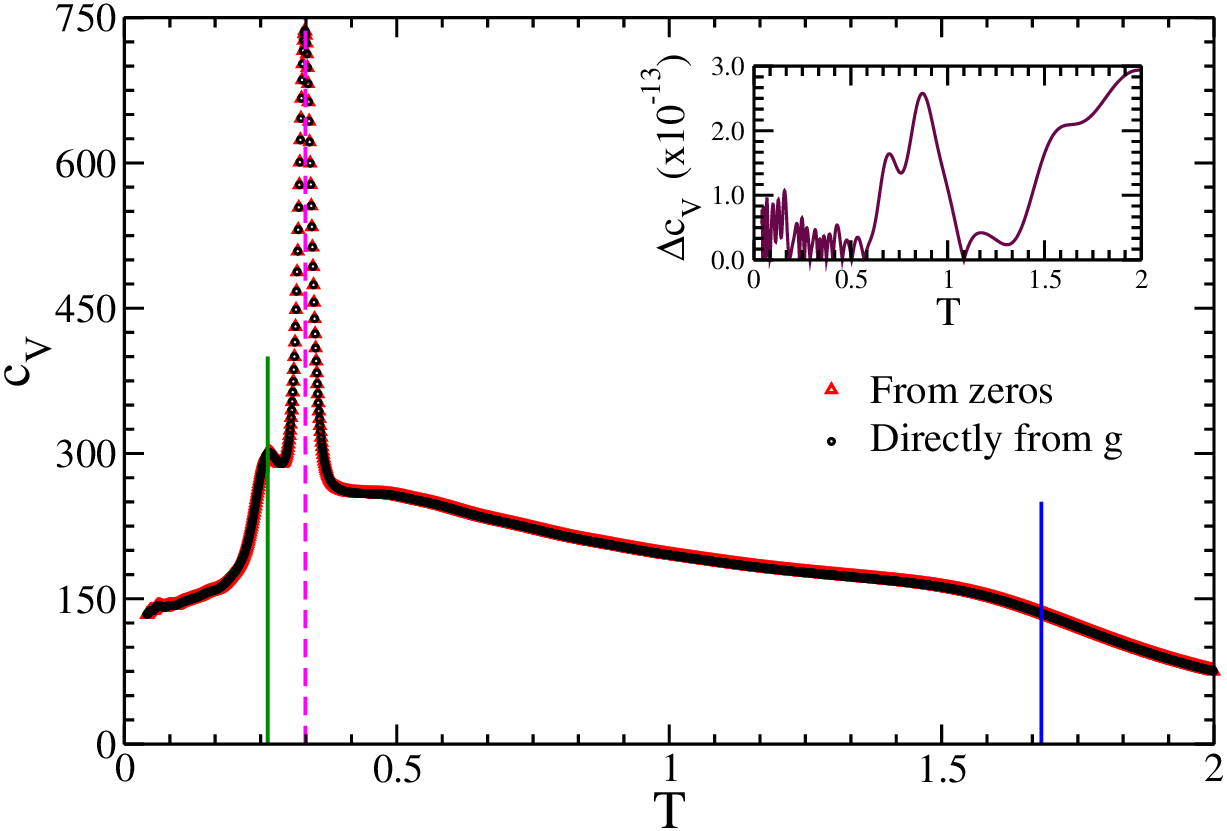}
& &
    \includegraphics[height=5.5cm,width=8.5cm]{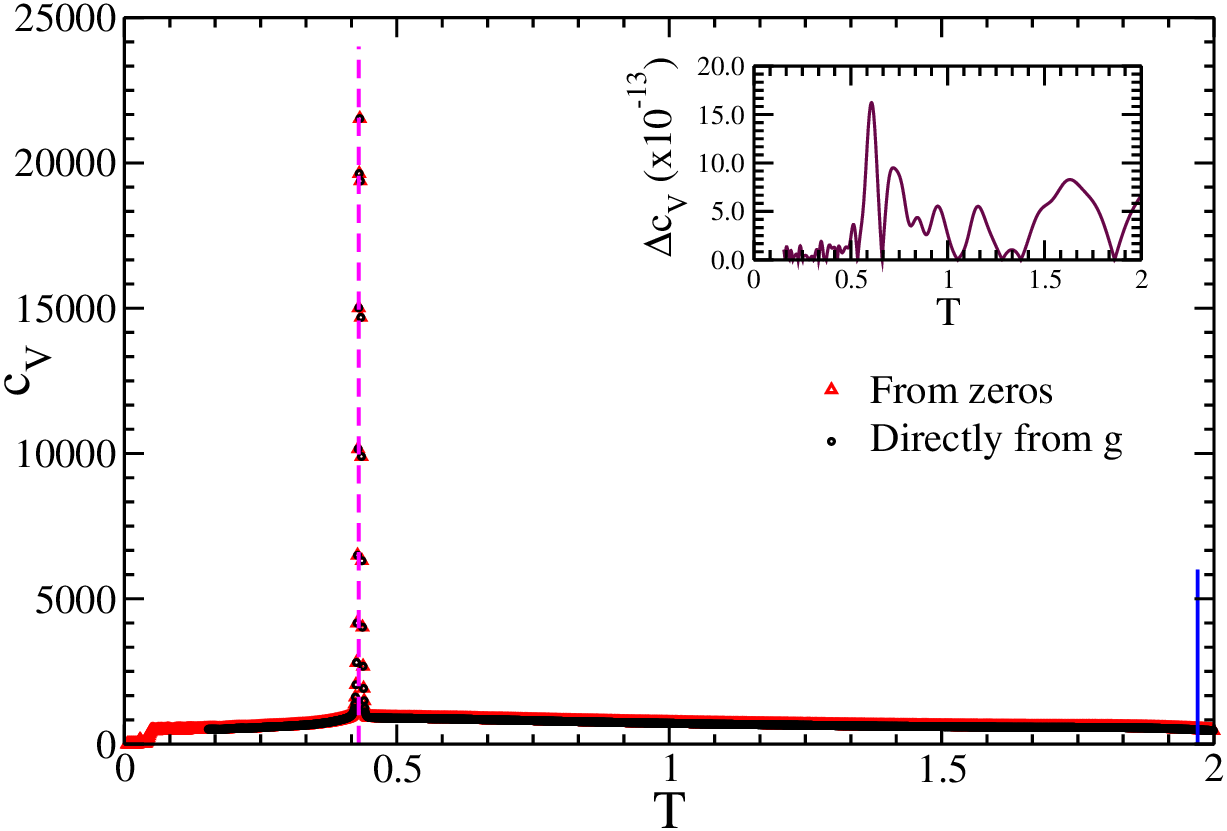}
\\
     (c) & & (d)
  \end{tabular}
  \caption{Heat capacity curves for chain sizes: (a) $L=35$, (b) $L=55$, (c)
$L=90$,
and (d) $L=300$. Plotted are the curves obtained from the zeros of the
partition function and, for comparison, 
by direct calculation from the density of states. The
inset shows the relative differences between
them. The small deviations make it evident that all zeros were identified
correctly. The vertical lines are located at the 
transition temperatures 
calculated from the leading zeros. Dashed and solid lines represent
first- and second-order-like
transitions, respectively. \label{specific_heat}}
\end{figure*}
An alternative approach to unravel transition properties of finite-size
systems is
the direct microcanonical analysis~\cite{gross1} of caloric quantities derived
from the entropy $S(E) = k_\mathrm{B}\ln{g(E)}$. The basic idea is that the
interplay of energy and entropy and, in particular, changes of it, signal
cooperative system behavior that can be interpreted as a transition (and in
the thermodynamic limit as a phase transition) of the system. Then
first and higher derivatives of $S(E)$ reveal the transition points of the
system in energy space. However, since the first derivative is the reciprocal
microcanonical temperature,
\begin{equation}
\beta(E) \equiv T^{-1}(E)  = \left( \frac{\partial S(E)}{\partial E}
\right)_{N,V},
\end{equation}
energetic transition points can also be associated with transition
temperatures. Transitions occur, if $\beta(E)$ responds least sensitively to
changes in the energy. The slope of the corresponding inflection points can
be used to distinguish first- and second-order transitions 
systematically. If
\begin{equation}
\gamma(E) = \left( \frac{\partial \beta(E)}{\partial E} \right)_{N,V} 
= \left( \frac{\partial^2 S(E)}{\partial E^2} \right)_{N,V}
\end{equation}
exhibits a positive-valued peak at the inflection point, the transition
resembles a first-order transition, whereas a negative-valued peak indicates
a second-order transition. This method is called microcanonical
inflection-point analysis~\cite{sslb1}. In the following, we will compare the
transition temperatures obtained from the leading zeros with microcanonical
estimates. 
%
\section{Results and discussion \label{results}}
%
Based on the density of states estimates obtained in multicanonical
simulations, we calculated the partition function zeros for the elastic
flexible polymer model for chain lengths $L$
ranging from $13$ to $309$ monomers. The structural transition behavior was
investigated previously by conventional canonical statistical analysis of
``peaks'' and ``shoulders'' of fluctuating energetic and structural
quantities as functions of the canonical temperature~\cite{svbj1,sbj1}.
Subsequently, the densities of states of this set of polymers were analyzed
systematically by means of microcanonical inflection-point
analysis, with particular focus on the typically hardly
accessible low-temperature transition behavior (freezing,
solid-solid transitions)~\cite{sslb1}. The microcanonical analysis is
based on estimates of the microcanonical entropy and \emph{its
derivatives}, and therefore requires highly accurate data. Therefore,
it is not only
interesting from the statistical physics point-of-view to study
the partition function zeros, but also for practical purposes. The
major information about structural transitions is already encoded in the
corresponding leading zeros which are rather simple to identify. The  
partition function zero method thus turns out to be a robust method
for the identification of transition points. It is, therefore, highly
interesting to verify whether this method is capable of finding indications
for the same transitions that have already been identified by means of
microcanonical inflection-point analysis.

Figure~\ref{root_map} shows the distributions of the zeros identified from the
discretized densities of states for specific chain lengths $L=35,55,90,$ and
$300$ and using the energy bin sizes $\varepsilon=0.07,0.11,0.20$, and
$0.29$,
respectively. It is worth noting that the zeros, and thus their distribution,
do
generally depend on the choice of $\varepsilon$, but the transition temperature 
estimates remain widely unaffected if $\varepsilon$ is changed.
Moreover,
since the data series used for the estimation of the density of states are
finite, different simulation runs yield different values of the zeros. 

Note that we plot the zeros differently than Ref.~\cite{taylor1}. In
our 
case they are strictly confined within a circle with radius 1 (the boundary at 
1 corresponds to infinite temperature). We also define the transition 
temperature differently for a finite system.  Ref.~\cite{taylor1} considers 
only the real part of the leading zero, whereas we prefer the absolute value, 
motivated by the fact that at first-order transitions the zeros lie on a circle 
whose radius is a unique estimator for the transition temperature.

The section of the map for $L=35$ shown
in Fig.~\ref{error_zoom} contains sets 
of zeros obtained in two independent simulations
(circles and triangles). By standard jackknife error
analysis~\cite{jack,efron1,wj2002,landaubinder1,mbbook1}, the
statistical error of the components of the complex zeros was estimated from
ten independent simulations and error bars are shown for the leading zeros
(squares) only (if larger than symbol size). Thus, for the analysis of
transitions, the method is sufficiently robust and enables the identification
of transition points.

We only analyze here the zero maps for $L=35,55,90,$ and
$300$, because these system sizes are
representative for the various transition behaviors that have been
systematically and uniquely identified for polymer chains with lengths in
the above mentioned interval in canonical~\cite{svbj1,sbj1,seaton1} and
microcanonical analyses~\cite{sslb1}. From these studies it is
known that in this model polymers with ``magic'' length
$L=13,55,147,309,\ldots$ possess a second-order-like collapse
(``gas-liquid'') transition and a very strong first-order-like freezing or
``liquid-solid'' transition from the compact, globular liquid phase into an
almost perfect icosahedral Mackay structure~\cite{mackay1}, where the facets
are arranged as fcc overlayers. For intermediate chain lengths, the optimal
packing in the solid phase can be Mackay or anti-Mackay (hcp overlayers),
depending on the system size and the temperature. In other words, for certain
groups of chain lengths, an additional ``solid-solid'' transition can be
found, in which anti-Mackay overlayers turn into energetically more preferred
Mackay facets at very low temperatures~\cite{sslb1,svbj1,sbj1,seaton1}.
This behavior of finite particle systems is also well known from atomic
clusters~\cite{northby1,wales1,doye2,frant1}. 

For the systems explicitly
discussed here, this means that we expect to find three transitions for
$L=35$ and $90$, whereas the solid-solid transition is absent for
$L=55$. For $L=300$, the liquid-solid and the solid-solid transition
merge and occur at about the same temperature. These transitions can be
distinguished microcanonically, but not canonically. Therefore, we do not
expect to find indications of separate transitions in the analysis of the
leading zeros. 

As earlier analyses revealed~\cite{svbj1,sslb1}, the
liquid-solid and solid-solid transitions for system sizes $31\le L\le 54$
have peculiar characteristics. Except for the special case
$L=38$ that forms a truncated fcc octahedron, these polymers crystallize in
two different ways by cooling down from
the liquid phase~\cite{svbj1}. With high
probability, more than one
icosahedral nucleus crystallizes out of the liquid by forming anti-Mackay
overlayers and by an additional solid-solid transition turns into a
single icosahedral nucleus with 13 monomers and a Mackay overlayer formed by
the remaining ones. Alternatively, with lower probability, the
anti-Mackay multi-core structure can also form out of the liquid via an
intermediate unstable phase dominated by a single-core structure with Mackay
overlayer. Therefore, the anti-Mackay solid phase is a mixed phase that also
contains Mackay morphologies. Therefore the liquid-solid transition for
these system sizes does not exhibit the same characteristic as for larger
polymers and is actually second-order-like~\cite{sslb1}. To conclude,
all three structural transitions for $L=35$ are second-order-like. The
corresponding zero maps shown in Figs.~\ref{root_map}(a)
and~\ref{error_zoom} indeed reveal three separate pairs of leading zeros that
represent these transitions.

The polymer chain containing 55 monomers is ``magic''. For this reason, it
exhibits a particularly strong liquid-solid transition at $T\approx 0.33$ into
a
perfect icosahedral conformation~\cite{svbj1} with complete Mackay overlayer.
A stable anti-Mackay phase does not exist and, therefore, no solid-solid
transition occurs. Consequently, the zero map shown in
Fig.~\ref{root_map}(b) reveals only two sets of leading zeros representing
the $\Theta$ collapse and the nucleation transition. The most striking
feature is the observation that there is an increased accumulation of zeros
on a circle that contains the pair of the leading zeros associated with the
liquid-solid transition. The circular distribution has to be attributed to 
the self-reciprocity of the partition function polynomial~\cite{wchen1} at a
phase transition with coexisting phases in which case the energetic canonical
distribution is bimodal and virtually symmetric. Therefore, the circular
pattern can be interpreted as the signature of first-order-like transitions in
the map of Fisher partition function zeros.

For the polymer with $L=90$ monomers, the structural transitions can clearly
be identified in the corresponding zeros map [Fig.~\ref{root_map}(c)]. The
liquid-solid transition into the anti-Mackay solid phase is represented by a
circular zeros distribution, but neither the collapse transition nor the
solid-solid crossover to icosahedral Mackay structures exhibit obvious
features in the zero distribution other than prominent locations of the
leading zeros. In correspondence with the previous microcanonical analysis,
these transitions are classified as of second order. It is worth mentioning
that the chain length $L=90$ is close to the threshold length ($L\approx
110$), at which in the canonical interpretation the liquid turns directly to
solid Mackay structures at the liquid-solid transition point and liquid-solid
and solid-solid transitions merge. 

No separate solid-solid transition occurs
for chain lengths $L>110$ until the next ``magic'' limit $L=147$ is
reached~\cite{sslb1,sbj1}, i.e., the Mackay phase is the only stable solid
phase. Microcanonically speaking, the solid-solid transition lies
energetically within the latent heat interval of the first-order liquid-solid
transition and can no longer be resolved in the canonical analysis
(the
specific heat exhibits only one sharp peak in these cases~\cite{sbj1}). The
zeros map shown in Fig.~\ref{root_map}(c) reveals a very pronounced circular
distribution and the projected intersection point with the positive $x$ axis
corresponds indeed to the liquid-solid transition temperature.

While $L=90$ is a length \emph{below} the anti-Mackay--Mackay threshold, our
last example $L=300$, is above the corresponding threshold in the following
segment of chain lengths that lies between two magic lengths, $147<L\le 309$
($L=309$ is the next ``magic'' chain length). The most surprising feature is
that in temperature space liquid-solid and solid-solid transitions merge,
whereas energetically both can be distinguished clearly as
first-order-like transitions~\cite{sslb1}. The trend is that 
the solid-solid transition will shift to \emph{higher} microcanonical
temperatures than the liquid-solid transition when increasing $L$ towards
$L=309$. This microcanonical crossover behavior has already been known in
other systems and is a pure finite-size effect~\cite{gnvb1}. The
corresponding root map shown in Fig.~\ref{root_map}(d) displays only the
general canonical behavior; therefore, only one circle represents this
first-order-like double-transition. 
\begin{table*}
\caption{Comparison of transition temperatures for solid-solid
(ss), liquid-solid (ls), and gas-liquid (gl) transitions for
$L=35,55,90,$
and $300$ as obtained by the partition function zero method ($T_\mathrm{z}$)
and by
microcanonical inflection-point analysis ($T_\mathrm{m}$). These estimates
are compared to peak positions
of the heat-capacity curves ($T_{c_V}^{\mathrm{ss,ls}}$) and fluctuations of
the radius of gyration ($T_{d\langle R\rangle/dT}^\mathrm{gl}$), respectively.
The maximum $1\sigma$ tolerance of
all estimates is $\pm 1$ in the last digit. There is no solid-solid transition
for the 55-mer. The solid-solid transition of the 300-mer can only be
distinguished from the liquid-solid transition in the microcanonical
inflection-point analysis.
\label{tab1}}
\setlength{\tabcolsep}{2.0mm}
\begin{tabular}{|c|ccc|ccc|ccc|} \hline
& \multicolumn{3}{|c|}{solid-solid} &
\multicolumn{3}{|c|}{liquid-solid} &
\multicolumn{3}{|c|}{gas-liquid}\\
$L$ & $T_\mathrm{z}^\mathrm{ss}$ & $T_\mathrm{m}^\mathrm{ss}$ &
$T_{C_V}^\mathrm{ss}$ &
$T_\mathrm{z}^\mathrm{ls}$ &
$T_\mathrm{m}^\mathrm{ls}$ &
$T_{C_V}^\mathrm{ls}$ &
$T_\mathrm{z}^\mathrm{gl}$ &
$T_\mathrm{m}^\mathrm{gl}$ & $T_{d\langle R\rangle/dT}^\mathrm{gl}$ \\ \hline
$35$ & $0.15$ & $0.14$ & $0.14$ & $0.39$ & $0.39$ & $0.38$ & $1.39$
& $1.39$ & $1.35$ \\
$55$ & N/A & N/A & N/A & $0.33$ & $0.33$ & $0.33$ & $1.53$ & $1.51$ &
$1.53$\\
$90$ & $0.26$ & $0.26$ & $0.27$ & $0.33$ & $0.33$ & $0.33$ & $1.68$
& $1.65$ & $1.67$ \\
$300$ & N/A & $0.44$ & N/A & $0.43$ & $0.43$ & $0.43$ & $1.97$ &
$1.88$ & $1.97$\\ \hline
\end{tabular}
\end{table*}

For the explicit estimation of the transition
temperatures from the Fisher zeros according to
Eq.~(\ref{transition_temperature}), there is the ambiguity to use either the
absolute values of the complex zeros or their real parts only,
\begin{equation}
T_\mathrm{tr} = -\frac{2\varepsilon}{k_\mathrm{B}\ln{(a_j^2  +
b_j^2)}} \approx -\frac{\varepsilon}{k_\mathrm{B}\ln{a_j}}.
\label{eq:trtemp}
\end{equation}
Both values differ for finite systems, but converge in the thermodynamic
limit. Since we already know that
distributions of zeros for first-order-like transitions are circular, we chose
to define
transition points by means of the absolute values (corresponding to the
radius of the circle). For the four examples that
we discuss here in more detail, the corresponding values are listed in
Table~\ref{tab1}. These estimates are in very good agreement with the
transition temperatures obtained by microcanonical analysis. Since the
$\Theta$ transition is only represented by a weak shoulder in the 
heat capacity curves shown in Fig.~\ref{specific_heat}, we consider in
these cases the corresponding peak positions of the fluctuations of the radius
of gyration, $d\langle R_\mathrm{gyr}\rangle/dT$ as a more appropriate
indicator of these transitions. This is a general problem of the canonical
analysis of fluctuating quantities and the major reason for the introduction
of methods that enable a unique identification of transition points even for
finite systems. 

For this reason both the zeros method and the microcanonical
inflection-point analysis are more useful for the definition of unique
transition temperatures than the conventional approach of the quantitative
analysis of fluctuating quantities. Furthermore, the analysis of zero
distributions or microcanonical inflection points allow the discrimination
between first- and second-order-like transitions. This information is not
easily accessible from ordinary canonical statistical analysis. In
Figs.~\ref{specific_heat}(a)-~\ref{specific_heat}(d), vertical lines are located at the positions
of the transition transitions obtained by the analysis of the Fisher zeros.  
\begin{figure}[b]   
\includegraphics[height=5.5cm,width=8.5cm]{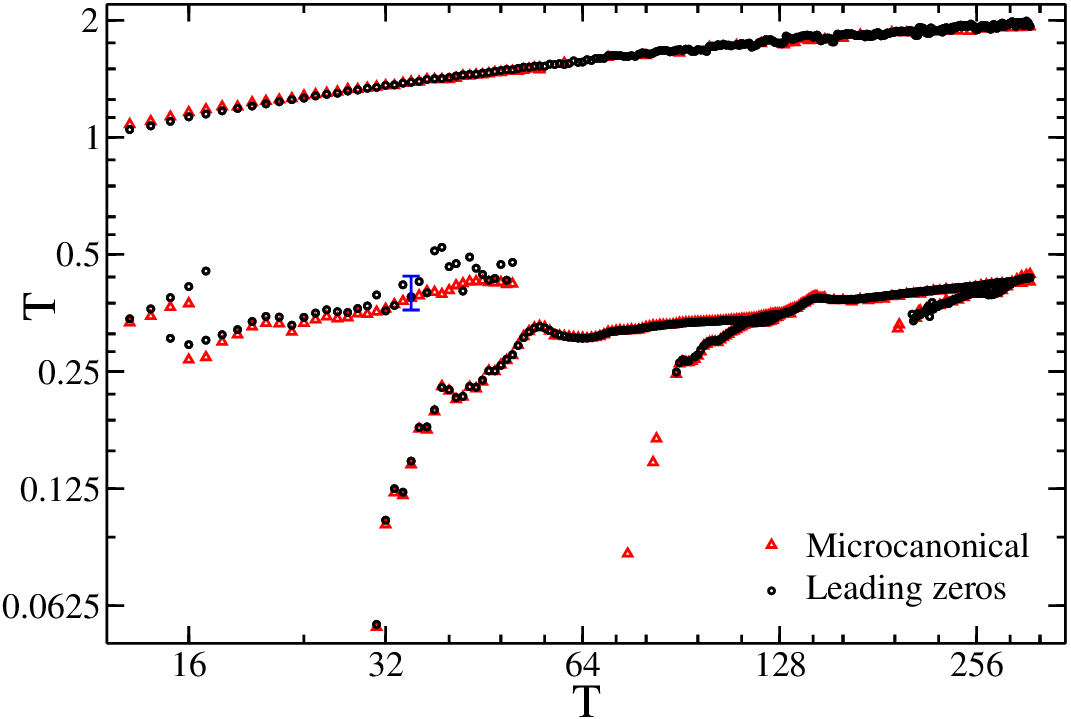}
  \caption{Transition temperatures of conformational transitions for
elastic, flexible
polymers with chain lengths ranging from $L=13$ to $309$. The black dots
represent the transition temperatures obtained from the leading zeros of the
partition function. For comparison, the transition temperatures obtained by
microcanonical inflection-point analysis are also shown (red triangles).
\label{temperature}}
\end{figure}

Fig.~\ref{temperature} summarizes our results of the Fisher zero
analysis for all chain lengths in the interval $13\le L \le 309$. For
comparison, the data from the microcanonical inflection-point analysis are
also shown. Although basically founded on the conventional canonical
understanding of temperature, the zeros method captures surprisingly many
details of transition behavior in finite polymer systems that were
formerly accessible only by microcanonical analysis. Note that the
temperature axis represents the microcanonical transition temperatures in the
case of the microcanonical analysis, whereas it scales canonical transition
temperatures obtained by the zeros method. These temperature estimates
do not typically coincide, and this is why larger deviations in the estimates
of transition temperatures seem to occur, particularly for small systems.
Furthermore, in those cases the indicators for the transitions are very weak
(which means that the transitions are also very weak) in both methods. This
explains why the numerical error of the transition temperatures is larger for
small systems than it is for larger ones ($L\ge 55$) that exhibit more stable
structural phases.
%
\section{Summary \label{summary}}
%
We calculated Fisher partition function zeros for a generic model of flexible,
elastic polymers on the basis of accurate estimates of the densities of states
for chain lengths $13\le L\le 309$. For there entire range of
chain lengths, we estimated transition temperatures systematically by
analyzing the leading zeros and their distributions. We identified the
gas-liquid and liquid-solid transition points, as well as the notoriously
difficult to find solid-solid transitions, which are only surface effects
but nonetheless relevant for finite systems. Our estimates of transition
temperatures are in
very good agreement with formerly obtained results by microcanonical
inflection-point analysis for the same model~\cite{sslb1}. By comparison with
the microcanonical classification scheme, we found numerical evidence for the
circular pattern of the zeros associated with coexisting states in
first-order-like transitions. Because the
zeros method is capable of revealing these signals, we conclude on the basis
of the results of our study that this method can be used for the
identification of transitions in small systems as well; otherwise it
would have had to be abandoned for this purpose.

We find that both the microcanonical
inflection-point analysis and the Fisher zeros method enable a quantitative
analysis of all ``transitions'' of a finite system. Both methods strongly
outperform the conventional canonical
approach of analyzing the ``peak-and-shoulder'' characteristics of
thermodynamic quantities such as the specific heat or canonical fluctuations
of order parameters as functions of the heat bath temperature. Whereas
microcanonical analysis
enables a more fine-tuned understanding of an individual transition (such as
the composition of subphase transitions), the zeros method is very robust and
the leading zeros are less sensitive to numerical errors. This
remarkable robustness can be attributed to the fact that the leading zeros
alone govern in all thermodynamic quantities the ultimate approach to the
transition point in the scale-free, universal regime. Numerical errors
can be interpreted as perturbations of the model, but such effective model
details have hardly any impact on the thermodynamic behavior of the system
near a transition point, even for relatively small systems. This is not the
case within the phases, where the location of the zeros depends more
sensitively on details. 
\begin{acknowledgments}
We would like to thank Shan-ho Tsai, Lucas A.~S.\ M\'ol, and Bismarck V.\
Costa for inspiring discussions.
JCSR thanks CNPq (National
Council for Scientific and Technological Development, Brazil) for the support
under grant No.\ PDE 202122/2011-5.
MB acknowledges partial support of this project by
the NSF under
Grant No.\ DMR-1207437 and by CNPq
under Grant No.\ 402091/2012-4.
\end{acknowledgments}
%
%


\begin{thebibliography}{99}
\bibitem{rem2}
In a monic polynomial of degree $n$, the $n$th coefficient
is unity, i.e., $g_n = 1$.
\end{thebibliography}
\end{document}